\documentclass[11pt]{article}
\usepackage{amsmath,amssymb,color,graphics,epsfig}

\usepackage{amsfonts}

\newcommand{\be}{\begin{equation}}
\newcommand{\ee}{\end{equation}}
\newcommand{\bea}{\setlength\arraycolsep{2pt} \begin{eqnarray}}
\newcommand{\eea}{\end{eqnarray}}
\newcommand{\nn}{\nonumber}

\def\ft#1#2{{\textstyle{\frac{\scriptstyle #1}{\scriptstyle #2} } }}
\def\fft#1#2{{\frac{#1}{#2}}}

\def\0{{\sst{(0)}}}
\def\1{{\sst{(1)}}}
\def\2{{\sst{(2)}}}
\def\3{{\sst{(3)}}}
\def\4{{\sst{(4)}}}
\def\5{{\sst{(5)}}}
\def\6{{\sst{(6)}}}
\def\7{{\sst{(7)}}}
\def\8{{\sst{(8)}}}
\def\sst#1{{\scriptscriptstyle #1}}

\begin{document}

\begin{flushright}
\end{flushright}

\vspace{25pt}
\begin{center}
{\large {\bf Black Holes with Vector Hair}}

\vspace{10pt}
 Zhong-Ying Fan

\vspace{10pt}
{\it Center for High Energy Physics, \\}
{\it Peking University, No.5 Yiheyuan Rd, Beijing 100871, P. R. China\\}

\vspace{40pt}

\underline{ABSTRACT}
\end{center}
In this paper, we consider Einstein gravity coupled to a vector field, either minimally or non-minimally, together with a vector potential of the type
$V=2\Lambda_0+\ft 12 m^2 A^2+\gamma_4 A^4$. For a simpler non-minimally coupled theory with $\Lambda_0=m=\gamma_4=0$, we obtain both extremal and non-extremal
black hole solutions that are asymptotic to Minkowski space-times. We study the global properties of the solutions and derive the first law of thermodynamics
using Wald formalism. We find that the thermodynamical first law of the extremal black holes is modified by a
one form associated with the vector field. In particular, due to the existence of the non-minimal coupling, the vector forms thermodynamic conjugates with the graviton mode
and partly contributes to the one form modifying the first law.
For a minimally coupled theory with $\Lambda_0\neq 0$, we also obtain one class of asymptotically flat extremal black hole solutions in general dimensions.
This is possible because the parameters $(m^2,\gamma_4)$ take certain values such that $V=0$. In particular,
we find that the vector also forms thermodynamic conjugates with the graviton mode and contributes to the corresponding first law, although the non-minimal coupling
has been turned off. Thus all the extremal black hole
solutions that we obtain provide highly non-trivial examples how the first law of thermodynamics can be modified by a either minimally or non-minimally coupled vector field.
 We also study Gauss-Bonnet gravity non-minimally coupled to a vector and
 obtain asymptotically flat black holes and Lifshitz black holes.

\vfill {\footnotesize  Email: fanzhy@pku.edu.cn\,.}

\thispagestyle{empty}

\pagebreak

\tableofcontents
\addtocontents{toc}{\protect\setcounter{tocdepth}{2}}




\section{Introduction}
In asymptotically Minkowski space-times, there exists a variety of no-hair theorems, which exclude the existence of black holes with hairs
(for recent reviews, see \cite{Herdeiro:2015waa,Volkov:2016ehx}). However, according to recent studies in the literature, the no-hair theorems are easily evaded. By numerical analysis,
it was demonstrated in \cite{Bizon:1990sr,Volkov:1989fi,kunzle:1990} that colored black holes with SU(2) Yang-Mills hairs do exist, although the solutions carry no global Yang-Mills charges.
The results were generalized to include rotation in
\cite{Kleihaus:2000kg,Kleihaus:2002ee}. Numerical
studies also indicate that rotating black holes with scalar hair \cite{Herdeiro:2014goa,Herdeiro:2015tia} and Proca hair \cite{Herdeiro:2016tmi} exist in asymptotically Minkowski space-times.
It was established also by numerical analysis in \cite{Lu:2015cqa,Lu:2015psa} that non-Schwarzschild black holes do exist in higher derivative gravity in four dimensions.

It is worth pointing out that generally the numerical analysis in the literature requires a delicate fine tuning. One may not expect the existence of such solutions
\footnote{However, as was shown in the numerical solutions, some of the hairy black holes may be achieved dynamically. More evidence can be found in \cite{Sanchis-Gual:2015lje}.}. Thus, it is deserved to find
exact black hole solutions with hairs. In fact, people have analytically
constructed many scalar hairy black holes that are asymptotic to Minkowski space-times \cite{Anabalon:2013qua,Anabalon:2012dw,Gonzalez:2013aca,Feng:2013tza,Fan:2015oca}
in Einstein gravity minimally coupled to a scalar field. By contrast, there are few analytical black hole solutions with vector hair. The only known examples are in \cite{Geng:2015kvs,Chagoya:2016aar},
 the authors obtain asymptotically flat black hole solutions by introducing a Horndeski-like non-minimal coupling between the gravity and the vector.
 Some more discussions on the existence of solutions with vector hair can be found in \cite{DeFelice:2016cri}.

The purpose of current paper is to construct new black hole solutions with vector hair. It was shown in \cite{Liu:2014tra} that for the general two-parameter family black hole solutions in Einstein-Proca
gravity, the first law of thermodynamics can be modified by a one form associated with the Proca\footnote{In \cite{Liu:2014tra}, this is explicitly demonstrated for the AdS black hole
solutions in Einstein-Proca gravity.
The analysis also holds when the vector field has self-interactions and can be straightforwardly generalized to asymptotically flat black holes.}. This is straightforwardly confirmed by numerical solutions.
However, people have not found any analytical examples so far. Hence, it is interesting to construct exact black hole solutions
whose first law of thermodynamics is modified by the vector hair.

We consider Einstein gravity either minimally or non-minimally coupled to a
generalized Proca field, together with a vector potential $V=2\Lambda_0+\ft 12 m^2 A^2+\gamma_4 A^4$. We obtain many new classes of asymptotically flat black hole solutions.
For a non-minimally coupled vector with $\Lambda_0=m=\gamma_4=0$, we obtain both extremal and non-extremal black hole solutions.
For a minimally coupled vector with $\Lambda_0\neq 0$, we also obtain one class of extremal black hole solutions.
This is possible because the parameters are such that $V=0$. In particular, using Wald formalism, we find that for all the extremal black holes,
the first law of thermodynamics is indeed modified by a one form associated with the vector field. Thus the solutions provide first analytical examples how the
thermodynamical first
law can be modified by a either minimally or non-minimally coupled vector field. By adding a Gauss-Bonnet term in the Lagrangian, we obtain more black hole solutions,
including asymptotically flat black holes and Lifshitz black holes.

The paper is organized as follows. In section 2, we discuss the Einstein gravity non-minimally coupled to a vector field. We derive the equations of motion and
analyze the structures of general static solutions. In section 3, we adopt Wald formalism to derive the first law for the general static solutions.
In section 4, we study a simpler non-minimally coupled theory and obtain asymptotically flat black hole solutions. We find that the first law
of the extremal black holes is modified by the vector. In section 5, we obtain extremal black hole solutions for a minimally coupled vector. The corresponding
first law is also modified by the vector. In section 6, we
introduce a Gauss-Bonnet term in the Lagrangian and obtain more black hole solutions. We conclude this paper in section 7.

\section{Einstein-Vector gravity}\label{sec2}

\subsection{Eoms and emerging gauge symmetry}
We consider Einstein gravity non-minimally coupled to a vector field, together with a generic vector potential (By ``vector potential",
I mean the potential term $V$ in the Lagrangian. It should not be confused with the vector field $A_\mu$). The Lagrangian density is given by
\be\label{genela}
\mathcal{L}= R-\beta A^2 R-\fft{1}{4}F^2-V(\psi) \,,
\ee
where $F=dA$, $\psi\equiv A_{\mu}A^{\mu}$ and $\beta$ is a constant characterizing the coupling strength between the vector and the curvature.
Note that the effective Newton's constant now becomes space-time dependent. It is inversely proportional to
\be \kappa(A)=1-\beta A^2  \,.\ee
To avoid ghost-like graviton mode, we shall require $\kappa(A)$ being positive definite throughout this paper.
The covariant equations of motions are
\be\label{eq1}
 G_{\mu\nu}=T_{\mu\nu}^{\rm (min)}+T_{\mu\nu}^{\rm (non)}\,,\qquad
\triangledown_\mu F^{\mu\nu} = 2A^\nu\Big(\beta R + \fft{dV}{d\psi}\Big)\,,
\ee
where $ G_{\mu\nu}=R_{\mu\nu}-\fft 12 R g_{\mu\nu}$ is the Einstein tensor. The energy-momentum tensors are given by
\bea
T_{\mu\nu}^{\rm (min)}&=&\fft{1}{2}\Big(F^2_{\mu\nu}-\fft 14 g_{\mu\nu}F^2  \Big)+\Big(\fft{dV}{d\psi}A_\mu A_\nu-\fft 12 g_{\mu\nu}V(\psi)\Big)\,,\nn\\
T_{\mu\nu}^{\rm (non)}&=&\beta\Big( A^2 G_{\mu\nu}+\big(g_{\mu\nu}\Box-\triangledown_\mu\triangledown_\nu\big)A^2+ R A_\mu A_\nu\Big)\,.
\label{eq2}\eea
In this paper, we consider the vector potential of the type
\be V=2\Lambda_0+\fft 12 m^2 A^2+\gamma_4 A^4 \,.\label{potential}\ee
The general theory is characterized by four independent parameters $(\beta\,,\Lambda_0\,,m^2\,,\gamma_4)$. The most simple solutions of the
theory are given by
\be G_{\mu\nu}=-\Lambda_0 g_{\mu\nu}\,,\qquad A=0 \,.\ee
It is clear that depending on the sign of the bare cosmological constant, the maximally symmetric vacuum is AdS ($\Lambda_0<0$), Minkowski ($\Lambda_0=0$) or
dS ($\Lambda_0>0$) space-times, respectively. Linearizing the equations of motion around the vacuum, we find that the linear fluctuations of the
equations are described by a massless graviton and a Proca with an effective mass
\be m^2_{\mathrm{eff}}=m^2+\ft{4 n}{n-2}\beta \Lambda_0\,,\ee
where $n$ denotes the space-time dimension. Note that the effective Proca mass can be generated in the vacuum in the presence of a non-minimal coupling even if the bare mass vanishes. The
U(1) gauge symmetry can emerge at the linear level when the parameters are such that the effective Proca mass vanishes $m^2_{\mathrm{eff}}=0$. In this case,
the vector field becomes simply a Maxwell field at the linear level.

Depending on the values of the parameters $(\beta\,,\Lambda_0\,,m^2\,,\gamma_4)$, there exist different theories, describing a either minimally or non-minimally coupled
vector. In this paper, we are interested in two special cases. The first case is a non-minimally coupled vector, described by
\be\label{simla}
\mathcal{L}= R-\beta A^2 R-\fft{1}{4}F^2 \,.
\ee
This corresponds to a vanishing vector potential with $\Lambda_0=m^2=\gamma_4=0$. The maximally symmetric vacuum is the Minkowski space-time. The gauge symmetry is restored at the linear level in any Ricci-flat metric, including the Minkowski vacuum, Schwarzschild and Kerr black holes. Interestingly, we find that the theory also admits
the solutions of Minkowski space-times with a constant vector
\be ds^2=-dt^2+dr^2+r^2 d\Omega_{n-2}^2\,,\qquad A=q_1 dt \label{lorentz}\,,\ee
where $d\Omega_{n-2}^2$ denotes the metric of unit $(n-2)$-sphere. Note that the solutions (\ref{lorentz}) have broken the Lorentz symmetry because the vector field, unlike the Maxwell field, is as physical as the field strength.
 As will be shown later, this plays an important role in deriving the thermodynamical first law of
asymptotically flat black hole solutions. Note that we obtain both extremal and non-extremal black hole solutions in sec.\ref{sec4}.

The second case we consider is
\be\label{simla2}
\mathcal{L}= R-2\Lambda_0-\fft{1}{4}F^2-\fft{1}{2}m^2 A^2-\gamma_4 A^4\,,
\ee
which describes a minimally coupled vector with self-interactions. With a non-vanishing bare mass $m\neq 0$, the gauge symmetry is broken and cannot be
restored in any background space-time. For $m=0$, the gauge symmetry can be restored at the linear level in any Einstein metric with the cosmological constant
$\Lambda_0$. For the theory with $\Lambda_0\neq 0$, the maximally symmetric vacuum is (A)dS space-times.
However, the theory also admits the solutions (\ref{lorentz}) when the parameters are such that the vector potential $V=0$. This is also true when the vector
contains higher order self-interactions. Substituting the solutions (\ref{lorentz}) into the equations of motion, we find
\be m^2=\fft{8\Lambda_0}{q_1^2}\,,\qquad \gamma_4=\fft{2\Lambda_0}{q_1^4} \,.\label{special2}\ee
It is straightforward to verify that the vector potential indeed vanishes.
Note that under the constraint, we obtain one class of asymptotically flat black holes in sec.\ref{sec5}.

\subsection{The structure of general static solutions}
In this paper, we focus on discussing the static solutions with spherical/toric/hyperbolic isometries. The most general ansatz is given by
\be\label{ansatz} ds^2=-h(r)dt^2+\fft{dr^2}{f(r)}+r^2d\Omega_{n-2,k}^2 \,,\qquad A=A_t(r)dt\,,\ee
where $d\Omega_{n-2,k}^2$ denotes the metric of the $(n-2)$ dimensional space with constant curvature $k=0\,,\pm 1$.
 We do not expect to find the general solutions of the equations of motion (\ref{eq1}-\ref{eq2}). Nevertheless, the properties of the general static solutions can be
discussed by developing Taylor series at asymptotic infinity and in the near horizon region. For later convenience, we focus on discussing the solutions that
are asymptotic to Minkowski space-times in the following
\footnote{The structure of the general static solutions that are asymptotic to AdS space-times and the corresponding first law have been well studied in \cite{Liu:2014tra}. }.

In the near horizon region, the metric functions and the vector can be expanded as
\bea
&&h=h_1(r-r_0)+h_2(r-r_0)^2+h_3(r-r_0)^3+\cdots\,,\nn\\
&&f=f_1(r-r_0)+f_2(r-r_0)^2+f_3(r-r_0)^3+\cdots\,,\nn\\
&&A_t=a_1(r-r_0)+a_2(r-r_0)^2+a_3(r-r_0)^3+\cdots\,.
\label{horizon}\eea
Note that the finite norm condition of the vector requires $A_t$ vanishes in the near horizon limit. Substituting the expansions into the equations of motion,
we find that there are three independent parameters in the near horizon region which we may take to be $(r_0\,,a_1\,,h_1)$. All the rest coefficients can be solved in terms of these three parameters. Note that $h_1$ is a trivial parameter, which is associated with the scaling symmetry of the time coordinate. Thus the near horizon solutions are characterized by two non-trivial parameters $(r_0\,,a_1)$.

At asymptotical infinity, the structure of the general static solutions strongly depends on the parameters of the vector potential. We shall consider the two special examples, corresponding to the Lagrangian density (\ref{simla}) and (\ref{simla2}), respectively. For the theory described by (\ref{simla}), we find that the asymptotic solutions contain either three independent parameters or only one independent parameter.
Let us first discuss the three parameters case. To leading order, we find
\be A_t=q_1-\fft{q_2}{r^{n-3}}+\cdots\,,\qquad h=1-\fft{\mu}{r^{n-3}}+\cdots\,,\qquad f=1-\fft{\tilde{\mu}}{r^{n-3}}+\cdots \,,\label{infinity1}\ee
where $\tilde{\mu}$ is determined by
\be \tilde{\mu}=\fft{(1-\beta q_1^2)\mu+4\beta q_1 q_2}{1+\beta q_1^2} \,.\ee
It should be emphasized that the asymptotic solutions (\ref{infinity1}) are obtained by solving the linearized equations of motion around the Minkowski space-times with a constant vector (\ref{lorentz}).
The fall-off mode $1/r^{n-3}$ in the metric functions is associated with the usually massless graviton mode. By plugging the expansions into the equations of motion,
we find that there are three independent parameters at infinity which we may take to be $(q_1\,,q_2\,,\mu)$. All the rest higher order coefficients can be determined
 by these three parameters. However, the boundary conditions on the horizon provide an algebraic constraint for the three parameters at asymptotic infinity.
 Consequently, when integrating out to infinity we find that the three parameters at infinity are determined by functions of the two non-trivial parameters of the near horizon solutions, namely
\be q_1=q_1(r_0\,,a_1)\,,\qquad q_2=q_2(r_0\,,a_1)\,,\qquad \mu=\mu(r_0\,,a_1) \,.\label{match}\ee
Equivalently, we may take the parametric relations by saying
\be q_2=q_2(\mu\,,q_1) \,.\label{parametric}\ee
Thus the general static solutions are characterized by two independent parameters.

It turns out that there exists another class of
solutions which contains only one independent parameter at infinity. To leading order, we find
\be A_t=q_1-\fft{q_2}{r^\sigma}+\cdots\,,\qquad h=1-\fft{\mu}{r^\sigma}+\cdots\,,\qquad f=1-\fft{\tilde{\mu}}{r^\sigma}+\cdots  \,,\label{infinity2}\ee
where $\sigma$ is a positive constant and $\sigma\neq n-3$. By substituting into the equations of motion, we find that
\be q_2=\fft{2\beta q_1(n-1)(1+\beta q_1^2)\mu}{(2n-5)\beta q_1^2-1}\,,\qquad \tilde{\mu}=\fft{\sigma(1+3\beta q_1^2)\mu}{(2n-5)\beta q_1^2-1}\,, \label{unusualpara}\ee
and $q_1$ has been fixed by a function\footnote{We find $q_1^2=C/D$, where $C=(1-4\beta)n+4(\beta-1)\pm 2\sqrt{(n-1)(2\beta+1)\Big( (2\beta-1)n-2\beta+3 \Big)}$
and $D=\beta\Big( (8\beta-5)n-2(4\beta-7)\Big) $.} of $n$ and $\beta$ (the detail is irrelevant in our discussion). It is clear that in this case there is only
one independent parameter at asymptotic infinity which we may take to be $\mu$. In addition, the parametric relation (\ref{parametric}) has been fixed by solving the
linearized equations of motion at asymptotic infinity. Hence, one may worry about the existence of this type solutions since it needs a delicate fine tuning of the
parameters on the horizon. Interestingly, we do obtain the black hole solutions
of this type with $\sigma=(n-3)/2$ in sec.\ref{sec41}. The key point we shall emphasize here is that the unconventional fall-off mode $1/r^\sigma$ in the metric functions
signifies the existence of longitudinal graviton modes, which are excited by the background vector. The general
analysis is presented in the Appendix.

For the theory described by (\ref{simla2}), the existence of asymptotically flat solutions in general requires a vanishing bare cosmological constant. However, for
non-vanishing $\Lambda_0$, the solutions can also exist provided the parametric relation (\ref{special2}), together with $\lim_{r\rightarrow \infty}A_t(r)=q_1 $.
The asymptotic solutions are of the type (\ref{infinity1}) with $\tilde{\mu}=\mu\,,q_2=(\mu q_1)/2$. For later convenience, we write the results in the following
\be A_t=q_1-\fft{\mu q_1}{2r^{n-3}}+\cdots\,,\qquad h=1-\fft{\mu}{r^{n-3}}+\cdots\,,\qquad f=1-\fft{\mu}{r^{n-3}}+\cdots \,.\label{infinity3}\ee

\section{Wald formalism and the first law of thermodynamics}
Wald formalism provides a systematic procedure for the derivation of first law of thermodynamics for a generic gravity theory. It was first developed
by Wald in \cite{wald1,wald2}. Variation of the action with respect to the metric and the matter fields, one finds
\be \delta \Big(\sqrt{-g}\mathcal{L} \Big)=\sqrt{-g}\Big(E_\phi \delta \phi+\nabla_\mu J^{\mu}\Big) \,,\ee
where $\phi$ collectively denotes the dynamical fields and $E_\phi=0$ are the equations of motion. For our gravity model (\ref{genela}), the current $J^\mu$ receives
contributions from both the gravity and the vector.
We find
\bea
&&J^\mu=J_{(G)}^\mu+J_{(A)}^\mu\,,\qquad J^\mu_{(G)}=G^{\mu\nu\rho\sigma}\nabla_\nu \delta g_{\rho\sigma}    \,,\nn\\
&&J^\mu_{(A)}=-F^{\mu\nu}\, \delta A_\nu+\beta G^{\mu\nu\rho\sigma}\big(\nabla_\nu A^2-A^2\nabla_\nu \big)\delta g_{\rho\sigma}\,,
\eea
where $G^{\mu\nu\rho\sigma}$ is the Wheeler-Dewitt metric, defined by
\be G^{\mu\nu\rho\sigma}=\frac 12 (g^{\mu\rho}g^{\nu\sigma}+g^{\mu\sigma}g^{\nu\rho})-g^{\mu\nu}g^{\rho\sigma}\,.\ee
Note that we have put the current associated with the non-minimally coupled term into the vector sector. For a given current $J^\mu$,  one can define a current 1-form
 and its Hodge dual
\be
J_\1=J_\mu dx^\mu\,,\qquad
\Theta_{\sst{(n-1)}}={*J_\1}\,.
\ee
When the variation is generated by an infinitesimal diffeomorphism $\xi^\mu=\delta x^\mu$, one can define an associated Noether current $(n-1)$-form as
\be
J_{\sst{(n-1)}}=\Theta_{\sst{(n-1)}}-i_\xi\cdot {*\mathcal{L}}\,,
\ee
where $i_\xi\cdot$ denotes the contraction of $\xi$ with the first index of the $n$-form ${}^*\mathcal{L}$ it acted upon. It is easy to show that the Noether
current $J_{(n-1)}$ is closed once the equations of motion are satisfied, namely
\be dJ_{(n-1)}=\mathrm{e.o.m} \,,\ee
where e.o.m denotes the terms proportional to the equations of motion. Thus one can further define a charge $(n-2)$-form as
\be J_{(n-1)}=dQ_{(n-2)} \,.\ee
It was shown in \cite{wald1,wald2} that when $\xi$ is a Killing vector, the variation of the Hamiltonian with respect to the integration constants of a specific solution is
given by
\be
\delta H=\frac{1}{16\pi}\Big[\delta \int_{\mathcal{C}}J_{\sst{(n-1})}- \int_{\mathcal{C}}d(i_\xi\cdot \Theta_{\sst{(n-2)}})\Big]=\frac{1}{16\pi}\int_{\Sigma_{n-2}}\Big[\delta Q_{\sst{(n-2)}}-i_\xi\cdot \Theta_{\sst{(n-2)}}\Big]\,.\label{generalwald}
\ee
where $\mathcal{C}$ is a Cauchy surface, $\Sigma_{n-2}$ is its two boundaries, one on the horizon and the other at infinity. For our vector-tensor theory,
the various quantities in the Wald formula can be straightforwardly derived. For the pure gravity, we have \cite{wald2}
\bea
J_{\sst{(n-1)}}^{(G)} &=&-2\varepsilon_{\mu c_1...c_{n-1}} \nabla_\nu\Big(\nabla^{[\mu}\xi^{\nu]}
\Big)  \,,\cr
Q_{\sst{(n-2)}}^{(G)} &=& -\varepsilon_{\mu\nu c_1...c_{n-2}}\,\nabla^{\mu}\xi^{\nu}\,,\cr
i_{\xi}\cdot \Theta_{(n-1)}^{(G)}&=&\varepsilon_{\mu\nu c_1...c_{n-2}}\xi^{\nu}\Big(
G^{\mu\lambda\rho\sigma} \nabla_\lambda \delta g_{\rho\sigma}\Big)\,.\label{various}
\eea
For the vector sector (with the non-minimally coupled term), we obtain
\bea
&&J^{(A)}_{(n-1)}=2\varepsilon_{\mu c_1\cdots c_{n-1}}\nabla_\nu\Big(-\ft 12 F^{\mu\nu}A^\sigma\xi_\sigma+\beta \big(A^2 \nabla^{[\mu}\xi^{\nu]}+2\xi^{[\mu}\nabla^{\nu]}A^2 \big)\Big) \,,\nn\\
&&  Q_{(n-2)}^{(A)}=\varepsilon_{\mu\nu c_1\cdots c_{n-2}}\Big(-\ft 12 F^{\mu\nu}A^\sigma\xi_\sigma+\beta \big(A^2 \nabla^{\mu}\xi^\nu+2\xi^\mu \nabla^\nu A^2\big) \Big)\,, \nn\\
&& i_\xi\cdot \Theta_{(n-1)}^{(A)}=\varepsilon_{\mu\nu c_1\cdots c_{n-2}}\xi^\nu\Big(-F^{\mu\nu}\, \delta A_\nu
+\beta\, G^{\mu\nu\rho\sigma}\big(\nabla_\nu A^2-A^2\nabla_\nu \big)\delta g_{\rho\sigma} \Big) \,.
\eea
Note that the Wald formula does not explicitly depend on the non-derivative terms of the Lagrangian. For $\beta=0$, the various quantities have been
given in \cite{Liu:2014tra,Liu:2014dva,Fan:2014ixa}.

Now we evaluate $\delta H$ for the general static solutions (\ref{ansatz}). Let $\xi=\partial/\partial t$, we obtain
\bea\label{waldformula1}
&&\delta H=\delta H^{(G)}+\delta H^{(A)}\,\nn\\
&&\delta H^{(G)}= \fft{\omega}{16\pi} r^{n-2}\, \sqrt{\fft{h}{f}}\, \Big(-\fft{n-2}{r} \Big)\delta f\,,
\eea
and
\bea\label{waldformula2}
&&\delta H^{(A)}=\delta H^{(A)}_{(\mathrm{min})}+\delta H^{(A)}_{(\mathrm{non})}\,,\nn\\
&&\delta H^{(A)}_{(\mathrm{min})}=-\frac{\omega}{16\pi}r^{n-2}\sqrt{\frac{h}{f}}\, \Big(\frac{f}{h}A_t\delta A_t'+\frac 12 A_t A_t'\big(\frac{\delta f}{h}-\frac{f\delta h}{h^2}\big)\Big)\,,\nn\\
&&\delta H^{(A)}_{(\mathrm{non})}=\fft{\beta\omega}{16\pi}r^{n-2}\sqrt{\fft{f}{h}}\Big(\fft{6h'}{h}A_t\delta A_t-4\delta(A_t A'_t)+A_t^2 \Delta(r)\Big)\,,\nn\\
&& \Delta(r)=\fft{2\delta h'}{h}+\Big(\fft{4A_t'}{A_t}-\fft{5h'}{h}\Big)\fft{\delta h}{h}+\Big(\fft{h'}{h}-\fft{2A_t'}{A_t}-\fft{n-2}{r} \Big)\fft{\delta f}{f}\,.
\eea
where $\omega$ is the volume factor of the $(n-2)$ dimensional space. By plugging the near horizon solutions (\ref{horizon}) into the Wald formula, it is easy to verify that
\be \delta H_+=T \delta S \,,\ee
is satisfied on the horizon. Here the temperature and entropy are given by
\be T=\fft{1}{4\pi}\sqrt{h'(r_0)f'(r_0)}\,,\qquad S=\fft{1}{4}\omega r_0^{n-2}\Big(1+\beta \fft{A^2_t(r_0)}{h(r_0)} \Big) \,.\ee
Note that the entropy is simply one quarter of the horizon area for non-extremal solutions whilst for extremal solutions the entropy is given by
\be S_{\mathrm{ext}}=\fft{1}{4}\omega r_0^{n-2}\Big(1+2\beta \fft{A'^2_t(r_0)}{h''(r_0)} \Big) \,.\ee
Evaluating $\delta H$ at both infinity and on the event horizon yields
\be \delta H_\infty=\delta H_+ \,.\ee
Thus the thermodynamical first law is simply
\be \delta H_\infty=T \delta S \,.\label{waldeq}\ee
Now we derive the thermodynamical first law for the general two-parameter family black hole solutions of the theories described by
the Lagrangian density (\ref{simla}) and (\ref{simla2}) respectively. Substituting the asymptotic solutions (\ref{infinity1}) into the Wald formula, we obtain
\bea\label{deltah1}
\delta H_\infty=\delta M&-&\fft{\omega}{16\pi}\Big( (n-3-4\beta)q_1\delta q_2+\ft{4\beta\big((2n-5)\beta q_1^2-1 \big)}{1+\beta q_1^2}q_2\delta q_1 \Big)\nn\\
                        &+&\fft{\beta\omega}{16\pi}\Big((n-4)q_1^2\delta \mu+\ft{2q_1\big(3\beta q_1^2(n-3)+n-5\big)}{1+\beta q_1^2}\mu \delta q_1  \Big)\,,
\eea
where $M$ is defined by
\be M\equiv \fft{(n-2)\omega}{16\pi}\mu \,.\ee
Note that $M$ is the ADM mass for Einstein gravity minimally coupled to matter fields. However, this is in general not true for non-minimally coupled theories. Nevertheless, we still refer
to the quantity $M$ as ``mass", or more strictly ``thermodynamic mass" \cite{Lu:2014maa,Liu:2015tqa}. The non-integrability
of $\delta H_\infty$ in (\ref{deltah1}) may be interpreted as that the solution has no well defined mass. Yet, we prefer the viewpoint proposed in \cite{Lu:2014maa,Liu:2015tqa}
that it is more instructive to interpret the relation (\ref{deltah1}) as providing a definition of the ``thermodynamic mass".
Note that the last bracket term in the equation (\ref{deltah1}) implies that the vector forms thermodynamic conjugates with the massless graviton mode
in the presence of non-minimal coupling. A similar phenomenon was first observed in Gauss-Bonnet gravity non-minimally coupled to a scalar for some critical
coupling constants \cite{Chen:2016qks}.
For vanishing $\beta$, the equation (\ref{deltah1}) reduces to the standard result of Einstein-Maxwell theories. Owing to (\ref{waldeq}), the thermodynamical first
law for the general static solutions can be expressed as
\bea \label{firstlaw1}
dM=T dS&+&\fft{\omega}{16\pi}\Big( (n-3-4\beta)q_1 dq_2+\ft{4\beta\big((2n-5)\beta q_1^2-1 \big)}{1+\beta q_1^2}q_2 dq_1 \Big)\nn\\
&-&\fft{\beta\omega}{16\pi}\Big((n-4)q_1^2 d\mu+\ft{2q_1\big(3\beta q_1^2(n-3)+n-5\big)}{1+\beta q_1^2}\mu  d q_1  \Big)\,.
\eea
For the asymptotic solutions (\ref{infinity3}) of the theory (\ref{simla2}), we obtain
\be \delta H_\infty=\delta M-\fft{(n-3)\omega}{32\pi}\Big( q_1^2\delta \mu+\fft 12 \mu \delta (q_1^2) \Big) \,.\label{deltah2}\ee
It is interesting to note that the vector also forms thermodynamic conjugates with the graviton mode even if the non-minimal coupling term has been turned off.
The corresponding first law can be written as
\be dM=T dS+\fft{(n-3)\omega}{32\pi}\Big( q_1^2 d\mu+\fft 12 \mu d (q_1^2) \Big) \,.\label{firstlaw2}\ee
In the following sections, we will also consider the solutions with electric charges of a Maxwell field $\mathcal{L}_{\mathcal{A}}=-\ft 14 \mathcal{F}^2$, where
$\mathcal{F}=d \mathcal{A}$. The first law becomes
\be dM=T dS+\Phi_e dQ_e+K(d q_1\,,d q_2\,,d \mu)\,, \label{genefirstlaw}\ee
where $K(d q_1\,,d q_2\,,d \mu)$ collectively denotes the one form associated with the vector in (\ref{firstlaw1}) and (\ref{firstlaw2}); $Q_e$
and $\Phi_e$ are the total electric charge and the conjugate potential defined by
\be Q_e=\fft{1}{16\pi}\int_{\Sigma_{n-2}}{}^*\mathcal{F} \,,\qquad \Phi_e=\mathcal{A}_t(\infty)-\mathcal{A}_t(r_0) \,.\label{maxwelldefinition}\ee

\section{With a non-minimally coupled vector field}\label{sec4}
In this section, we study the non-minimally coupled theory, described by (\ref{simla}). We obtain both non-extremal and extremal black hole solutions that
are asymptotical to Minkowski space-times. The extremal solutions can be generalized to include electric charges. We will also study a more general
non-minimally coupled theory and obtain extremal black hole solutions.

\subsection{Non-extremal black holes}\label{sec41}
There exists one class of non-extremal black hole solutions in general dimensions for the non-minimal coupling constant $\beta=(n-3)/\Big(2(n-1)\Big)$.
The solutions read
\bea && ds^2=-fdt^2+\fft{dr^2}{f}+r^2d\Omega_{n-2}^2\,, \nn\\
     && A=\sqrt{\ft{2(n-1)}{(n-3)}}f dt\,,\quad f=1-\fft{q^{\fft{n-3}{2}}}{r^{\fft{n-3}{2}}}\,,
\label{sol1}\eea
where $d\Omega_{n-2}^2$ denotes the metric of $(n-2)$ dimensional sphere with unit radius. Note that the solutions also exactly solve the linearized equations of
motion around the Minkowski space-time with a constant vector (\ref{lorentz}). It is easy to verify that the parametric relations (\ref{unusualpara}) are indeed
satisfied. Evaluating $\delta H_\infty$ yields
\be \delta H_\infty=\fft{(n-2)(n-3)\omega}{32\pi}q^{n-4}\delta q  \,.\ee
The black hole mass can be defined by $\delta M\equiv \delta H_\infty$, giving rise to
\be M=\fft{(n-2)\omega}{32\pi}q^{n-3}\,.\ee
The temperature and entropy are given by
\be T=\fft{n-3}{8\pi q}\,,\qquad S=\fft 14 \omega q^{n-2} \,. \ee
It follows that the thermodynamical first law
\be dM=T dS \,,\ee
straightforwardly holds. In addition, the mass, temperature and entropy satisfy a Smarr relation
\be M=\fft{n-2}{n-3}T S \,.\ee
It is worth pointing out that the solutions (\ref{sol1}) can be distinguished from the Schwarzschild black holes in the weak field limit in which the Newtonian concept of gravitational force is recovered. For example, in four dimension it has $1/r^{3/2}$-law rather than the usual $1/r^2$-law associated with the Schwarzschild black
holes. However, this does not mean the Newtonian gravity is excluded in our theory. In fact, as mentioned in sec.\ref{sec2}, Schwarzschild black holes are also the solutions of the theory. It is interesting that our theory now predicts the existence of new black holes with vector hair that have stronger gravitational force than the Schwarzschild black holes. The existence of such black holes provides a new candidate how the Newton's inverse-squared law can be modified and may be tested by observational data in the future.

\subsection{Extremal black holes}
The second class solution we find in the theory (\ref{simla}) is a extremal black hole, which is only valid in the $n=4$ dimension.

\subsubsection{Neutral black holes}
The black hole solution reads
\bea && ds^2=-fdt^2+\fft{dr^2}{f}+r^2d\Omega_2^2\,, \nn\\
    && A=\ft{2}{\sqrt{1-4\beta}}\sqrt{f} dt\,,\quad f=\Big(1-\fft{q}{r}\Big)^2\,.
\label{sol2}\eea
Reality of the solution requires $\beta<1/4$. Note that for $\beta=0$, the solution becomes an extremal Reissner-Nordstr\"{o}m (RN) black hole. For
generic $\beta$, the solution also looks like an extremal RN black hole. In fact, the background is Ricci flat, namely $R=0$ and hence the gauge symmetry of the vector emerges at the linear level. However, asymptotically the vector field $A_t(\infty)$ is a physical constant rather than a pure gauge. Consequently, the Lorentz symmetry of the space-time is breaking at asymptotic infinity. As emphasized earlier, this plays an important role in deriving the first law of thermodynamics. Thus the solution can be distinguished from the RN black hole via its global properties. This is also valid for other extremal black hole solutions in this paper.

The temperature vanishes and the entropy is given by
\be S=\fft{\omega q^2}{4(1-4\beta)} \,,\ee
which generally is no longer one quarter of the area of the horizon. Wald calculation shows $\delta H_\infty=0$, which is consistent with above results. However, the thermodynamical first law is not trivially satisfied. According to
the discussions of Eq.(\ref{firstlaw1}), the first law of extremal black holes reads
\be
dM=\fft{\omega}{16\pi}\Big( (1-4\beta)q_1 dq_2+\ft{4\beta(3\beta q_1^2-1 )}{1+\beta q_1^2}q_2 dq_1-\ft{2\beta q_1(3\beta q_1^2-1)}{1+\beta q_1^2}\mu  d q_1 \Big)\,.
\ee
Note that the $q_1^2 d\mu$ term in the general first law Eq.(\ref{firstlaw1}) is cancelled in the four dimension. Calculating the various quantities
\be M=\fft{\omega q}{4\pi}\,,\qquad \mu=2q\,,\qquad q_1=\fft{2}{\sqrt{1-4\beta}}\,,\qquad q_2=\fft{2q}{\sqrt{1-4\beta}} \,,\ee
it is easy to verify that the above first law is indeed satisfied.
 The mass and the ``vector charge" also satisfy a Smarr formula
\be M=\fft{\omega}{16\pi}(1-4\beta)q_1 q_2 \,.\ee

\subsubsection{Charged black holes}
Notice that since $q_1$ is not a free integration constant in the solution (\ref{sol2}), it may not be a sufficiently non-trivial example to test the first law
(\ref{firstlaw1}). To let $q_1$ free, we introduce an additional Maxwell field whose Lagrangian density is given by $\mathcal{L}_{\mathcal{A}}=-\fft 14 \mathcal{F}^2$, where $\mathcal{F}=d\mathcal{A}$. We find that above solution can be generalized to include electric charges. We obtain

\bea && ds^2=-fdt^2+\fft{dr^2}{f}+r^2d\Omega_2^2\,,\quad \mathcal{A}=-\fft{Q}{r}dt\,,\nn\\
    && A=q_1\sqrt{f} dt\,,\qquad f=\Big(1-\fft{q}{r}\Big)^2\,,
\label{sol3}\eea
where the charge parameter is given by
\be Q=q\sqrt{4+q_1^2(2\beta-1)}\,.\ee
Note that now $q_1$ is indeed a free integration constant and the solution contains two independent parameters which we may take to be $(q_1\,,q)$.
It is clear that the solution is still extremal $T=0$. The entropy is given by
\be S=\fft 14 \omega q^2(1+\beta q_1^2) \,.\ee
Various thermodynamic quantities can be computed as
\be M=\fft{\omega q}{4\pi}\,,\qquad q_2=q_1 q\,,\qquad \Phi_e=\sqrt{4+q_1^2(2\beta-1)}\,,\qquad Q_e=\fft{\omega Q}{16\pi}\,.\ee
It is straightforward to verify that the first law
\be dM=\Phi_e dQ_e+\fft{\omega}{16\pi}\Big( (1-4\beta)q_1 dq_2+\ft{4\beta(3\beta q_1^2-1 )}{1+\beta q_1^2}q_2 dq_1-
\ft{2\beta q_1(3\beta q_1^2-1)}{1+\beta q_1^2}\mu  d q_1 \Big) \,,\label{chargefirst}\ee
and the Smarr formula
\be M=\Phi_e Q_e+\fft{\omega}{16\pi}(1-4\beta)q_1q_2 \,,\ee
are satisfied. We remark that every term on the r.h.s of (\ref{chargefirst}) is non-vanishing because $q_1$ is a free parameter. Thus, the charged solution provides a more non-trivial example than the neutral solution to test the first law of thermodynamics.

\subsection{More solutions for general non-minimal coupling function}
Now we consider the theories with a generic non-minimal coupling function, described by
\be \mathcal{L}=R-(-1)^{s+1}\beta A^{2s}R-\fft 14 F^2 \,,\label{genenonla}\ee
where $s=1,2,3,\cdots$ is a positive integer. The theory (\ref{simla}) is included as $s=1$ case. It turns out that we can obtain more extremal black hole solutions
for generic $s$. We will also derive the first law of thermodynamics using Wald formalism. For the general static solutions (\ref{ansatz}), the total variation
of the Hamiltonian $\delta H$ is still given by (\ref{waldformula1}) and (\ref{waldformula2})
but now the term associated with the non-minimal coupling is replaced by
\bea
&&\delta H^{(A)}_{(\mathrm{non})}=\fft{\beta\omega}{16\pi}r^{n-2}\sqrt{\fft{f}{h}}\fft{A_t^{2s-2}}{h^{s-1}}
\Big(\fft{2s(2s+1)h'}{h}A_t\delta A_t-4s(2s-1)A'_t\delta A_t-4s A_t\delta A'_t+A_t^2 \Delta(r)\Big)\,,\nn\\
&& \Delta(r)=\fft{2s\delta h'}{h}+\Big(\fft{4s A_t'}{A_t}-\fft{(2s+3)h'}{h}\Big)\fft{s\delta h}{h}+\Big(\fft{s h'}{h}-\fft{2s A_t'}{A_t}-\fft{n-2}{r} \Big)\fft{\delta f}{f}\,.
\eea

\subsubsection{Neutral black holes}
In the $n=4$ dimension, we obtain an extremal black hole solution
\bea && ds^2=-fdt^2+\fft{dr^2}{f}+r^2d\Omega_2^2\,, \nn\\
    && A=q_1\sqrt{f} dt\,,\quad f=\Big(1-\fft{q}{r}\Big)^2\,,
\label{sol6}\eea
where $q_1$ has been fixed by the non-minimal coupling constant
\be 4\beta q_1^{2s}-q_1^2+4=0 \,.\label{q1beta}\ee
The temperature vanishes and the entropy is given by
\be S=\fft 14 \omega q^2\Big( 1+(-1)^{s+1}\beta q_1^2 \Big) \,.\label{entropy6}\ee
Since the solution contains only one independent integration constant, we need first analyze the structure of the general two-parameter family black hole solutions and
derive the first laws using Wald formalism. At asymptotic infinity, the vector and the metric functions behave as
\be A_t=q_1-\fft{q_2}{r}+\cdots\,,\qquad h=1-\fft{\mu}{r}+\cdots\,,\qquad f=1-\fft{\tilde{\mu}}{r}+\cdots \,,\ee
where
\be \tilde{\mu}=\fft{\big(1-(2s-1)\beta q_1^{2s} \big)\mu+4s\beta q_1^{2s-1}q_2}{1+\beta q_1^{2s}} \,.\ee
Plugging the expansions into the Wald formula, we obtain
\bea
\delta H_\infty=\delta M&-&\fft{\omega}{16\pi}\Big( (1-4s\beta q_1^{2s-2})q_1\delta q_2+\ft{4s\beta q_1^{2s-2}\big((2s+1)\beta q_1^{2s}-2s+1 \big)}{1+\beta q_1^{2s}}q_2\delta q_1 \Big)\nn\\
                        &-&\fft{\beta\omega}{16\pi}\Big(2(s-1)q_1^{2s}\delta \mu-\ft{2s q_1^{2s-1}\big((2s+1)\beta q_1^{2s}-2s+1\big)}{1+\beta q_1^{2s}}\mu \delta q_1  \Big)\,.
\eea
Thus for extremal solutions, the thermodynamical first law should be
\bea
dM&=&\fft{\omega}{16\pi}\Big( (1-4s\beta q_1^{2s-2})q_1 d q_2+\ft{4s\beta q_1^{2s-2}\big((2s+1)\beta q_1^{2s}-2s+1 \big)}{1+\beta q_1^{2s}}q_2 d q_1 \Big)\nn\\
&&+\fft{\beta\omega}{16\pi}\Big(2(s-1)q_1^{2s}d\mu-\ft{2s q_1^{2s-1}\big((2s+1)\beta q_1^{2s}-2s+1\big)}{1+\beta q_1^{2s}}\mu d q_1  \Big)\,.
\eea
Note that the $q_1^{2s} d\mu$ term appears for $s>1$ cases. The first law can be straightforwardly verified provided
\be M=\fft{\omega q}{4\pi}\,,\qquad \mu=2q\,,\qquad q_2=q_1 q\,, \ee
together with the parametric relation (\ref{q1beta}). The Smarr relation is given by
\be M=\fft{\omega}{16\pi}(1-4\beta q_1^{2s-2})q_1 q_2 \,.\ee

\subsubsection{Charged black holes}
The solution (\ref{sol6}) can be generalized to include electric charges
\bea && ds^2=-fdt^2+\fft{dr^2}{f}+r^2d\Omega_2^2\,,\quad \mathcal{A}=-\fft{Q}{r}dt\nn\\
    && A=q_1\sqrt{f} dt\,,\qquad f=\Big(1-\fft{q}{r}\Big)^2\,,
\label{sol7}\eea
where the charge parameter is given by
\be Q=q\sqrt{4(1+\beta q_1^{2s})-q_1^2} \,.\ee
Hence, the solution contains two independent integration constants. The temperature vanishes and the entropy is still given by (\ref{entropy6}). By simple calculations,
we obtain the various thermodynamic quantities
\be M=\fft{\omega q}{4\pi}\,,\qquad \mu=2q\,,\qquad q_2=q_1 q\,,\qquad \Phi_e=\sqrt{4(1+\beta q_1^{2s})-q_1^2}\,,\qquad Q_e=\fft{\omega Q}{16\pi} \,.\ee
It is easy to verify that the thermodynamical first law
 \bea
dM=\Phi_e dQ_e &+&\fft{\omega}{16\pi}\Big( (1-4s\beta q_1^{2s-2})q_1 d q_2+\ft{4s\beta q_1^{2s-2}\big((2s+1)\beta q_1^{2s}-2s+1 \big)}{1+\beta q_1^{2s}}q_2 d q_1 \Big)\nn\\
&+&\fft{\beta\omega}{16\pi}\Big(2(s-1)q_1^{2s}d\mu-\ft{2s q_1^{2s-1}\big((2s+1)\beta q_1^{2s}-2s+1\big)}{1+\beta q_1^{2s}}\mu d q_1  \Big)\,,
\eea
and the Smarr formula
\be M=\Phi_e Q_e+\fft{\omega}{16\pi}(1-4\beta q_1^{2s-2})q_1 q_2 \,,\ee
are satisfied.
\section{With a minimally coupled vector field}\label{sec5}

\subsection{Neutral black holes}
For the minimally coupled Einstein-Vector theory described by (\ref{simla2}), we obtain exact extremal black hole solutions in general dimensions
\bea && ds^2=-fdt^2+\fft{dr^2}{f}+r^2d\Omega_{n-2}^2\,, \nn\\
     && A=\sqrt{\ft{2(n-2)}{(n-3)}}\sqrt{f} dt\,,\quad f=\Big(1-\fft{q^{n-3}}{r^{n-3}}\Big)^2\,,
\label{sol4}\eea
with
\be m^2=\fft{4(n-3)}{(n-2)}\Lambda_0\,,\qquad \gamma_4=\fft{(n-3)^2}{2(n-2)^2}\Lambda_0 \,.\ee
The solutions enjoy an extremal RN-like form but now the gauge symmetry of the vector cannot be restored at the linear level due to a nonzero bare mass. The Lorentz symmetry of the space-time is also breaking at asymptotic infinity because the vector $A_t(\infty)$ is a non-vanishing physical parameter.

The temperature vanishes and the entropy is simply one quarter of the horizon area. According to Eq.(\ref{firstlaw2}),
the corresponding first law is given by
\be dM=\fft{(n-3)\omega}{32\pi}\Big( q_1^2 d\mu+\fft 12 \mu d (q_1^2) \Big) \,.\label{reducelaw}\ee
Provided the various quantities
\be M=\fft{(n-2)\omega}{8\pi}q^{n-3}\,,\qquad q_1=\sqrt{\fft{2(n-2)}{n-3}}\,,\qquad \mu=2q^{n-3} \,.\ee
it is straightforward to verify that the first law (\ref{reducelaw}) is indeed satisfied. Note that this unconventional first law is a consequence of
the breaking of Lorentz symmetry of the space-time at asymptotic infinity. In addition, there exists a Smarr-like relation
\be M=\fft{(n-3)\omega}{32\pi}\mu q_1^2 \,.\ee

\subsection{Charged black holes}
The extremal solutions (\ref{sol4}) can be straightforwardly generalized to include electric charges. We obtain
\bea && ds^2=-fdt^2+\fft{dr^2}{f}+r^2d\Omega_{n-2}^2\,, \quad \mathcal{A}=-\fft{Q}{r^{n-3}}dt\nn\\
     && A=q_1 \sqrt{f} dt\,,\qquad f=\Big(1-\fft{q^{n-3}}{r^{n-3}}\Big)^2\,.
\label{sol5}\eea
Some parameters are specified by
\be m^2=\fft{8\Lambda_0}{q_1^2}\,,\qquad \gamma_4=\fft{2\Lambda_0}{q_1^4}\,,\qquad Q=q^{n-3}\sqrt{\ft{2(n-2)}{(n-3)}-q_1^2} \,.\ee
Note that the charged solutions contain two independent integration constants. Reality of the solutions requires $q_1^2<2(n-2)/(n-3)$. The solutions are also extremal.
The entropy is still given by one quarter of the horizon area. By simple calculations, we obtain
\be M=\fft{(n-2)\omega}{8\pi}q^{n-3}\,,\qquad \mu=2q^{n-3}\,,\qquad \Phi_e=\sqrt{\ft{2(n-2)}{(n-3)}-q_1^2}\,,\qquad Q_e=\fft{(n-3)\omega Q}{16\pi}\,.\ee
It follows that the thermodynamical first law
\be dM=\Phi_e dQ_e+\fft{(n-3)\omega}{32\pi}\Big( q_1^2 d\mu+\fft 12 \mu d (q_1^2) \Big) \,,\ee
and the Smarr-like relation
\be M=\Phi_e Q_e+\fft{(n-3)\omega}{32\pi}\mu q_1^2 \,,\ee
are satisfied.
\section{Gauss-Bonnet black holes with vector hair}
Now we consider Gauss-Bonnet (GB) gravity non-minimally coupled to a vector field. The Lagrangian density is given by
\be \mathcal{L}=R-\beta A^2 R+\alpha \big(R^2-4R_{\mu\nu}^2+R_{\mu\nu\rho\sigma}^2\big)-\fft 14 F^2-V(\psi)\,. \label{GBlagrangian}\ee
where $\alpha$ is the Gauss-Bonnet coupling constant. The covariant equations of motion are
\be
 G_{\mu\nu}+\alpha H_{\mu\nu}=T_{\mu\nu}^{\rm (min)}+T_{\mu\nu}^{\rm (non)}\,,\qquad
\triangledown_\mu F^{\mu\nu} = 2A^\nu\Big(\beta R + \fft{dV}{d\psi}\Big)\,,\label{GBeom}
\ee
where
\be H_{\mu\nu}=2\Big(R R_{\mu\nu}-2R_{\mu\rho}R^\rho_{\nu}-2R^{\tau\sigma}R_{\tau\mu\sigma\nu}+R_{\mu\rho\tau\sigma}R_{\nu}^{\ \rho\tau\sigma} \Big)
               -\fft 12 g_{\mu\nu}\Big(R^2-4R_{\tau\sigma}^2+R_{\lambda\rho\tau\sigma}^2 \Big)\,,
               \ee
The energy-momentum tensors are still given by (\ref{eq2}). For GB gravity, the various quantities of the Wald formula have been explicitly given in \cite{Chen:2016qks,Fan:2014ala}. For the general
static solutions (\ref{ansatz}), we have
\be
\delta H^{(G)}= \fft{\omega}{16\pi} r^{n-2}\, \sqrt{\fft{h}{f}}\, \Big(
-\fft{n-2}{r} + \fft{2\alpha(n-2)(n-3)(n-4) (f-k)}{r^3} \Big)\delta f\,.
\label{GBwald}\ee
The total variation of the Hamiltonian $\delta H$ is still given by (\ref{waldformula1}) and (\ref{waldformula2})
with $\delta H^{(G)}$ replaced by (\ref{GBwald}). In addition, for non-extremal black holes with an event horizon $r_0$, the Wald entropy is given by
\be S=\fft{1}{4}\omega r_0^{n-2}\Big(1+\beta \fft{A^2_t(r_0)}{h(r_0)}+\fft{2\alpha k(n-2)(n-3)}{r_0^2} \Big) \,,\ee
whilst for extremal black holes
\be S_{\mathrm{ext}}=\fft{1}{4}\omega r_0^{n-2}\Big(1+2\beta \fft{A'^2_t(r_0)}{h''(r_0)}+\fft{2\alpha k(n-2)(n-3)}{r_0^2} \Big) \,.\ee
In order to obtain black hole solutions as many as possible, we will also introduce an additional Maxwell field
when necessary.

\subsection{Asymptotically flat black holes}
For vanishing vector potential $V=0$, we obtain a non-extremal black hole solution with electric charges in the $n=5$ dimension for $\beta=1/4$. The solution reads
\bea
&&ds^2=-fdt^2+\fft{dr^2}{f}+r^2d\Omega^2_3\,,\quad \mathcal{A}=-\fft{Q}{r}dt\,,\nn\\
&& A=2f dt\,,\qquad f=1-\fft{q}{r}\,,
\eea
where $Q^2=-6\alpha q^2$. Thus the reality of the solution requires $\alpha\leq 0$. Note that the solution contains only one independent parameter.
The temperature and entropy are given by
\be T=\fft{1}{4\pi q}\,,\qquad S=\fft 14 \omega q^3\Big(1+12\alpha q^{-2} \Big) \,.\ee
Evaluating $\delta H_\infty$ yields
\be \delta H_\infty=\fft{3\omega}{16\pi}q \delta q \,,\ee
implying that the black hole mass is given by
\be M=\fft{3\omega }{32\pi} q^2\,.\ee
The electric charge and the conjugate potential can be computed as
\be Q_e=\fft{\omega Q}{8\pi}\,,\qquad \Phi_e=\fft{\sqrt{-6\alpha}}{q} \,.\ee
It follows that the thermodynamical first law
\be dM=T dS+\Phi_e dQ_e \,,\ee
and the Smarr relation
\be M=\fft 32(T S+\Phi_e Q_e) \,,\ee
are satisfied.

\subsection{Lifshitz black holes}
It is known that in GB gravity, there are in general two distinct (A)dS vacuum solutions. However, for a critical GB coupling $\alpha=\fft{\ell^2}{2(n-3)(n-4)}$,
the two (A)dS vacua can coelesce into one. Here $\ell$ denotes the effective AdS radius. Recently, it is established in \cite{Chen2016} that
at the critical point the linearized equations of motion are exactly cancelled. Thus, the gravity has no graviton in the sense that
there is no propagator in the theory.
It was shown in \cite{Dehghani:2010kd,Dehghani:2010gn} that Lifshitz vacuum solutions\footnote{In fact, the most general static solutions
at the critical point contain two types of solutions
\bea
&&\mathrm{type\,1}:\quad -g_{tt}=g^{rr}=\fft{r^2}{\ell^2}+k-\fft{\mu}{r^{\fft{n-5}{2}}}\,,\nn\\
&&\mathrm{type\,2}:\quad g^{rr}=\fft{r^2}{\ell^2}+k\,,\qquad g_{tt}\mathrm{\,\,is\,\, an\,\, arbitrary\,\, function\,\, of\,\, r\,.\nn}
\eea
It is clear that the Lifshitz vacuum solutions are special examples of the type 2 solutions.} are allowed at the critical point.
In the presence of a generalized Proca field, we find that the Lifshitz solutions are allowed for generic GB coupling in our gravity model with the vector potential (\ref{potential}). We obtain
\be ds^2=\ell^2\Big(-r^{2z}dt^2+\fft{dr^2}{r^2}+r^2dx^i dx^i\Big)\,,\qquad A=p \ell r^z dt \,,\ee
 provided
\bea
&&p^2=\fft{2(z-1)\Big(1-2(n-3)(n-4)\alpha \ell^{-2} \Big)}{2\beta+(1-2\beta)z}\,,\nn\\
&&\gamma_4=\fft{1}{4p^2\ell^2}\Big [m^2\ell^2-(n-2)z-2\beta\Big(n^2+(2z-3)n+2(z-1)^2 \Big) \Big ]\,,\nn\\
&&\Lambda_0=-\ft {1}{2\ell^2}(n-1)(n-2)\Big(1-(n-3)(n-4)\alpha \ell^{-2}  \Big)-   \nn\\
&&\qquad\,\,  \ft {1}{4\ell^2}\Big[m^2\ell^2+z^2-2\beta \Big(n^2+(4z-3)n+2(2z^2-4z+1) \Big) \Big]p^2+\ft 32\gamma_4p^4  \,.
\eea
Note that there are still three parameters to be freely chosen. This allows us to construct exact Lifshitz black hole solutions.

\subsubsection{Neutral black holes}
We first obtain two classes of neutral black hole solutions.

\subsubsection{Case 1:}
The first class solution reads
\bea
&&ds^2=-r^{2z}\tilde{f}dt^2+\fft{dr^2}{r^2\tilde{f}}+r^2dx^idx^i\,,\nn\\
&&A=p r^z \tilde{f}dt\,,\qquad \tilde{f}=1-\fft{q^{n+z-2}}{r^{n+z-2}}\,,
\eea
where the effective AdS radius has been set to unity. Various parameters are specified by
\bea
&&p^2=\fft{2(z-1)(z-n)}{(z-2)(n+z-2)}\,,\quad \beta=\fft{(n-2)(z-2)}{2(n-z)(n+2z-2)}\,,\\
&&\alpha=\fft{(n-2)(z-1)(2n+3z-2)}{2(n-3)(n-4)(z-2)\Big(n^2+(3z-4)n+2(z-1)(z-2)  \Big)}\,,\nn\\
&&\Lambda_0=\fft{(n+z-2)\Big(n^2+2(z-2)n-2(z-1)(z^2-2z+2) \Big)}{2(z-2)(n+2z-2)}\,,\,\, m^2=\fft{(n-2)(z-2)(n+z-2)}{n-z}\,,\nn\\
&&\gamma_4=\fft{(z-2)(n+z-2)\Big(z n^3-(z+2)n^2-2(z^3-2z^2+4z-4)n+4(z-1)(z^2-z+2)  \Big)}{8(z-1)(n+2z-2)(n-z)^2}\nn\,.
\eea
Note that all the quantities are measured in the units of AdS radius. Since $p$ should be real, the Lifshitz exponent satisfies $1<z<2$ or $z>n$. Evaluating $\delta H_\infty$ yields
\be \delta H_\infty=\fft{(n-2)\omega}{16\pi}(n+z-2)q^{n+z-3}\delta q \,,\ee
implying that the black hole mass is given by
\be M=\fft{(n-2)\omega}{16\pi}q^{n+z-2} \,.\ee
Note that the mass is linearly proportional to the coefficient of the fall-off mode $1/r^{n+z-2}$. In fact, for AdS black holes ($z=1$) the mode coincides with the usually massless graviton mode. Thus it may be interpreted as the condensate of the massless graviton in asymptotically Lifshitz space-times (for a detailed discussion, see \cite{Liu:2014dva}).

The temperature and entropy are given by
\be T=\fft{(n+z-2)q^z}{4\pi}\,,\qquad S=\fft 14\omega q^{n-2} \,.\ee
It follows that the thermodynamical first law and the Smarr-like relation
\be dM=T dS\,,\qquad M=\fft{n-2}{n+z-2}T S \,,\label{liffirstlaw}\ee
straightforwardly hold. It is worth pointing out that for Lifshitz planar black holes, there exists an extra scaling symmetry
\be t\rightarrow \lambda^{-z}t\,,\qquad r\rightarrow \lambda r \,,\qquad x^i\rightarrow \lambda^{-1}x^i\,,\qquad ds^2\rightarrow ds^2 \,,\ee
which leads to above Smarr-like relation \cite{Liu:2015tqa,Hyun:2015tia}.
\subsubsection{Case 2:}
The second class solution reads
\bea
&&ds^2=-r^{2z}\tilde{f}dt^2+\fft{dr^2}{r^2\tilde{f}}+r^2dx^idx^i\,,\nn\\
&&A=p r^z \tilde{f}dt\,,\qquad \tilde{f}=1-\fft{q^{\ft 12(n+z-2)}}{r^{\ft 12(n+z-2)}}\,,
\eea
with various parameters specified by
\bea
&&p^2=\fft{8(z-1)^2}{(z-2)(n+z-2)}\,,\quad \beta=\fft{n+(z+1)(z-2)}{2(z-1)(n+3z-2)}\,,\\
&&\alpha=\fft{ (z-2)n^2-4(z-1)n-z(5z^2+2z-8) }{2(z-2)(n-3)(n-4)(n+z-2)(n+3z-2)}\,,\nn\\
&&\Lambda_0=-\fft{(n+z-2)\Big((z-2)n^2+(5z^2-12z+8)n+2(4z^3-13z^2+14z-4)  \Big)}{4(z-2)(n+3z-2)}\,,\nn\\
&&m^2=\fft{(n+z-2)\Big((z+2)n^2+(7z^2-6z-8)n+2(z-2)(4z^2-3z-2)  \Big)}{4(z-1)(n+3z-2)}\,,\nn\\
&&\gamma_4=\fft{(z-2)(n+z-2)\Big((z-2)n^3-2(3z-4)n^2-(5z^3+2z^2-16z+8)n+2z(5z^2+2z-8)  \Big)}{ 128(z-1)^3(n+3z-2) }\nn\,.
\eea
The reality condition requires $z>2$. Wald calculation shows
\be \delta H_\infty=\fft{(n-2)\omega}{32\pi}(n+z-2)q^{n+z-3}\delta q \,,\ee
implying that the black hole mass is given by
\be M=\fft{(n-2)\omega}{32\pi}q^{n+z-2} \,.\ee
It should be emphasized that the mass is proportional to the coefficient square of the fall-off mode $1/r^{\ft 12(n+z-2)}$, which clearly does not correspond to the usually (transverse and massless) graviton mode. A physical interpretation is the fall-off
mode is associated with the condensate of longitudinal graviton mode which is excited owing to the breaking of Lorentz symmetry in asymptotically Lifshitz space-time.

The temperature and entropy are given by
\be T=\fft{(n+z-2)q^z}{8\pi}\,,\qquad S=\fft 14\omega q^{n-2} \,.\ee
It follows that the thermodynamical first law and the Smarr-like relation (\ref{liffirstlaw}) are satisfied.

\subsubsection{Charged black holes}
Introducing an additional Maxwell field, we obtain two classes of charged Lifshitz black hole solutions.

\subsubsection{Case 1:}
The first class solution is valid for $z=3(n-2)$. The solutions read
\bea
&&ds^2=-r^{6(n-2)}\tilde{f}dt^2+\fft{dr^2}{r^2\tilde{f}}+r^2dx^idx^i\,,\quad A=p r^{3(n-2)} \tilde{f}dt\,,\nn\\
&&\mathcal{A}=\lambda q^{n-2}r^{2(n-2)}dt\,,\quad \quad \tilde{f}=1-\fft{ q^{2(n-2)}}{r^{2(n-2)}}\,,
\eea
where various parameters are specified by
\bea
&&\lambda^2=\fft{(3n-5)(11n-24)+20\alpha(n-2)(n-3)(n-4)(3n-8)}{2(n-2)(9n-16)}\,,\nn\\
&&p^2=\fft{(3n-7)\Big(7n-13-20\alpha(n-2)(n-3)(n-4) \Big)}{(n-2)(9n-16)}\,,\nn\\
&&\beta=\fft{(n-2)\Big(1-8\alpha(n-3)(n-4) \Big)}{2\Big(7n-13-20\alpha(n-2)(n-3)(n-4) \Big)}\,,\nn\\
&&\Lambda_0=-\fft{(n-2)}{2( 9n-16 )}\Big(99n^2-397n+394-4\alpha(n-2)(n-3)(n-4)(27n-58) \Big)\,,\nn\\
&&m^2=\fft{2(n-2)^2 \Big(23n-44-112\alpha(n-2)(n-3)(n-4) \Big)}{7n-13-20\alpha(n-2)(n-3)(n-4)}\,,\nn\\
&&\gamma_4=\fft{\alpha(n-3)(n-4)(9n-16)^2(n-2)^3 }{(3n-7) \Big(7n-13-20\alpha(n-2)(n-3)(n-4)  \Big)^2  }\,.
\eea
Since $\lambda$ and $p$ should be real, the GB coupling satisfies
\be -\fft{(3n-5)(11n-24)}{20(n-2)(n-3)(n-4)(3n-8)}<\alpha<\fft{7n-13}{20(n-2)(n-3)(n-4)} \,.\ee
For $\alpha=0$, the solutions are obtained in \cite{Geng:2015kvs}. Wald calculation implies that the black hole mass is given by
\be M=-\fft{\omega q^{4(n-2)}}{64\pi(9n-16)}\Big( (3n-7)(5n-8)+20\alpha(n-2)(n-3)(n-4)(3n-8) \Big) \,.\ee
By definition, the electric charge can be computed as
\be Q_e=\fft{(n-2)\omega}{8\pi}\lambda q^{n-2} \,,\ee
while the usual definition for the conjugate potential (\ref{maxwelldefinition}) becomes invalid because the gauge potential diverges at asymptotic infinity.
A refined definition was provided in  \cite{Liu:2014dva,Fan:2014ala,Fan:2015yza,Fan:2015aia}
\be \Phi_e\equiv \mathcal{A}^{\mathrm{reg}}_t(\infty)-\mathcal{A}_t(r_0)\,,\qquad
\mathcal{A}^{\mathrm{reg}}_t(\infty)=\mathcal{A}_t(\infty)-\mathcal{A}^{\mathrm{div}}_t(\infty)  \,,\ee
where $\mathcal{A}^{\mathrm{div}}_t(\infty)$ denotes the divergent term of $\mathcal{A}_t(\infty)$. For our solutions, we find
\be \Phi_e=-\lambda q^{3(n-2)} \,.\ee
The temperature and entropy are given by
\be T=\fft{n-2}{2\pi}q^{3(n-2)}\,,\qquad S=\fft 14\omega q^{n-2} \,.\ee
It follows that the first law of thermodynamics
\be dM=T dS+\Phi_e dQ_e \,,\ee
is satisfied. The Smarr-like relation is given by
\be M=\fft 14(T S+\Phi_e Q_e) \,.\ee

\subsubsection{Case 2:}
The second class solution is valid for generic $z$. The solutions read
\bea
&&ds^2=-r^{2z}\tilde{f}dt^2+\fft{dr^2}{r^2\tilde{f}}+r^2dx^idx^i\,,\quad A=p r^z \tilde{f}dt \,,\nn\\
&&\mathcal{A}=\sqrt{\ft{2(n-2)}{z-n+2}} q^{n-2}r^{z-n+2}dt\,,\qquad \tilde{f}=1-\fft{ q^{2(n-2)}}{r^{2(n-2)}}\,,
\eea
with
\bea
&&p^2=\fft{2(z-1)(z-n+1)}{(z-2)(n-2)  }\,,\qquad \beta=\fft{(n-2z)(2n-z-4)}{4(z-n+1)(2n+z-4)}\,, \nn\\
&&\alpha=\fft{(z-1)(5n-8)(2n-z-4)}{4(z-2)(n-2)(n-3)(n-4)(2n+z-4)} \,,\quad \Lambda_0=\fft{z(n-2)(n+3z-4)(2n-3z)}{2(z-2)(2n+z-4)}   \,,\nn\\
&&m^2=\fft{(n-2)\Big(4n^3-20n^2-(15z^2-10z-32)n+2(4z^3+11z^2-10z-8)\Big)}{ (z-n+1)(2n+z-4) }\,,\nn\\
&&\gamma_4=\fft{(z-2)(n-2) }{16(z-1)(2n+z-4)(n-z-1)^2 }\Big(6n^4+(5z-46)n^3-(28z^2+7z-128)n^2\nn\\
&&\qquad \quad+2(10z^3+46z^2-15z-76)n-4(z-1)(z+4)(z^2+6z+4)  \Big) \,.
\eea
The reality condition requires $1<z<2$ or $z>n-1$. Wald calculation shows $\delta H_\infty=0$. Thus the black hole mass vanishes. Other thermodynamic quantities are given by
\bea
&&T=\fft{n-2}{2\pi}q^z\,,\qquad S=\fft 14 \omega q^{n-2}\,,\nn\\
&&\Phi_e=-\sqrt{\ft{2(n-2)}{z-n+2}}\,q^z\,,\quad Q_e=\fft{\omega }{8\pi}\sqrt{\ft{z-n+2}{2(n-2)}}\,q^{n-2}\,.
\eea
It follows that the thermodynamical first law
\be 0=T dS+\Phi_e dQ_e \,,\ee
and the Smarr-like relation
\be 0=T S+(n-2)\Phi_e Q_e \,,\ee
are satisfied.

\section{Conclusion}
In this paper, we study Einstein gravity either minimally or non-minimally coupled to a vector field, together with a vector potential of the form $
V=2\Lambda_0+\ft 12 m^2A^2+\gamma_4 A^4$. In order to obtain exact black hole solutions as many as possible, we also introduce an additional Maxwell field and
a Gauss-Bonnet term in the Lagrangian when necessary.

For a non-minimally coupled vector with vanishing potential, corresponding to $\Lambda_0=m=\gamma_4=0$, we obtain both extremal and non-extremal black hole
solutions that are asymptotic to Minkowski space-times. We adopt Wald formalism to study the global properties of the solutions and derive the first law of thermodynamics. We find that the first law of the
extremal black holes is modified by a one form associated with the vector via a very sophisticated way. In general,
the one form contains two copies of contributions. On one hand, the two independent modes of the vector at asymptotic infinity,
one dimensionless and the other dimensionful, form thermodynamic conjugates, providing one copy of the contributions. On the other hand, owing to the existence of the non-minimal coupling,
the dimensionless mode of the vector can also form thermodynamic conjugates with the massless graviton mode, giving rise to the other copy of the contributions.

For a minimally coupled vector with non-vanishing bare cosmological constant $\Lambda_0\neq 0$, the maximally symmetric vacuum is (A)dS space-times. However,
the theories also admit the solutions of Minkowski space-times with a constant vector when the parameters are such that the vector potential vanishes $V=0$.
We obtain one class of extremal black hole solutions in general dimensions, that are asymptotic to Minkowski space-times. Interestingly, the vector also forms thermodynamic conjugates with the
graviton mode and contributes to the corresponding first law although the non-minimal coupling has been turned off.

Thus all the extremal black hole solutions that we obtain provide analytical examples how the first law of thermodynamics can be modified by a either minimally or non-minimally coupled vector field. With a Gauss-Bonnet term turned on, we obtain more black hole solutions, including asymptotically flat black holes and Lifshitz black holes.

Finally, to end this paper, we shall point out that there are a few examples of exact scalar hairy dynamical black holes having been found for a given Lagrangian
in the literature \cite{Zhang:2014sta,Lu:2014eta,Xu:2014xqa,Zhang:2014dfa,Fan:2015tua,Ayon-Beato:2015ada,Fan:2015ykb,Fan:2016yqv}.
The solutions provide analytical examples describing black holes formation, which is of great importance and interests in General Relativity.
Motivated by this, we are also trying to construct exact dynamical solutions with vector hair.
Unfortunately, we find that none of the solutions reported in this paper can become dynamic. This is left as an open problem.

\section*{Acknowledgments}
The author is grateful to Prof. Hong Lu and Prof. Bin Chen for valuable comments on the first version of the manuscript. This work was in part supported by NSFC Grants No.~11275010, No.~11335012 and No.~11325522.

\section{Appendix: Linear fluctuations around the Minkowski space-time with a constant Vector}
Now we analyze the linear fluctuations of the theory described by
\be
\mathcal{L}= R-\beta A^2 R-\fft{1}{4}F^2 \,,
\ee
around the Minkowski space-time with a constant vector
\be ds^2=-dt^2+dr^2+r^2 d\Omega_{n-2}^2 \,,\qquad A=q_1 dt\,. \label{lorentz2}\ee
The covariant equations of motion are given by (\ref{eq1}-\ref{eq2}) with the vector potential $V=0$. For later convenience, we write them in the following
\be
 G_{\mu\nu}=T_{\mu\nu}^{\rm (min)}+T_{\mu\nu}^{\rm (non)}\,,\qquad
\triangledown_\mu F^{\mu\nu} = 2 \beta R A^\nu\,,
\ee
where
\bea
&&G_{\mu\nu}=R_{\mu\nu}-\fft 12 R g_{\mu\nu}\,,\quad T_{\mu\nu}^{\rm (min)}=\fft{1}{2}\Big(F^2_{\mu\nu}-\fft 14 g_{\mu\nu}F^2  \Big)\,,\nn\\
&&T_{\mu\nu}^{\rm (non)}=\beta\Big( A^2 G_{\mu\nu}+\big(g_{\mu\nu}\Box-\triangledown_\mu\triangledown_\nu\big)A^2+ R A_\mu A_\nu\Big)\,.
\eea
Linearizing the metric and the vector as
\be g_{\mu\nu}=\eta_{\mu\nu}+h_{\mu\nu}\,,\qquad A_{\mu}=q_1\delta^0_\mu +A^L_{\mu}\,,\qquad F_{\mu\nu}=F^L_{\mu\nu} \,,\ee
we obtain the linearized equations of motion
\be (1+\beta q_1^2)\mathcal{G}^L_{\mu\nu}=\beta (\eta_{\mu\nu}\partial^2-\partial_\mu\partial_\nu)\psi^L+\beta q_1^2 \delta^0_\mu \delta^0_\nu R^L \,,\qquad
\partial_\mu F_L^{\mu\nu}=0 \,.\ee
Here $\mathcal{G}^L_{\mu\nu}=R^L_{\mu\nu}-\ft 12 \eta_{\mu\nu}R^L$ is the linearized Einstein tensor, $\partial^2=\eta^{\mu\nu}\partial_\mu \partial_\nu$,
$\psi^L=(g^{\mu\nu}A_\mu A_\nu)^L$ and $R^L=(g^{\mu\nu}R_{\mu\nu})^L$. Note that to obtain the linearized equations, we frequently use the fact that the background is
Ricci-flat, namely $\bar{R}=\bar{R}_{\mu\nu}=\bar{G}_{\mu\nu}=0$. It is clear that the vector becomes a Maxwell field at the linear level. With the non-minimal coupling turned off $\beta=0$, the linearized Einstein equations become $\mathcal{G}^L_{\mu\nu}=0$, as expected because
the gauge symmetry of the Maxwell field is restored at the full non-linear level and the background (\ref{lorentz2}) becomes the Minkowski vacuum. For a non-vanishing $\beta$,
the Lorentz symmetry of the background (\ref{lorentz2}) is effectively broken. Consequently, the longitudinal mode of the metric fluctuations is excited due to
the back-reaction effects of the vector. To be concrete, taking trace of the linearized equations, we obtain
\be \Big(n-2+(n-4)\beta q_1^2 \Big) R^L=-2\beta(n-1)\partial^2\psi^L \,,\ee
where $\psi^L$ can be written more explicitly $\psi^L=-(q_1^2 h_{00}+2 q_1 A^L_0)$. It is clear that the longitudinal graviton mode become dynamical due to the existence of
a source term from the vector. This is significantly different from the case in the Minkowski vacuum where one finds $R^L=0$ and hence the graviton is transverse
and massless.

Finally, we present some linearized curvatures in the following
\be R^L_{\mu\nu}=\fft 12\Big(\partial_{\sigma}\partial_\mu h^\sigma_\nu+\partial_\sigma \partial_\nu h^\sigma_\mu-\partial^2 h_{\mu\nu}
-\partial_\mu \partial_\nu h \Big)\,,\qquad R^L=-\partial^2 h+\partial_\mu \partial_\nu h^{\mu\nu} \,.\ee
Note that we cannot take any gauge conditions such as the transverse gauge because the background (\ref{lorentz2}) breaks the Lorentz symmetry.

\end{document}